\documentclass[]{spie}  

\usepackage{amsmath,amsfonts,amssymb}
\usepackage{graphicx}
\usepackage[colorlinks=true, allcolors=blue]{hyperref}
\usepackage{braket}
\usepackage{breqn}
\usepackage{booktabs}
\usepackage{dcolumn}
\usepackage{enumitem}
\usepackage[table]{xcolor}
\usepackage{blkarray}
\usepackage{bm}
\usepackage{float,titlesec}

\usepackage{amsmath}
\usepackage{algorithm}
\usepackage[noend]{algpseudocode}

\allowdisplaybreaks 

\definecolor{firstColor}{cmyk}{0.053333333,0.0088888888,0.0088888888,0.017777777}
\definecolor{secondColor}{cmyk}{0.01,0.01,0.11,0.02}

\title{ Optimization and synchronization of programmable quantum communication channels}

\author[a]{Mee Seong Im, Ph.D.}
\author[b]{Venkat R. Dasari, Ph.D.}
\affil[a]{U.S. Military Academy, West Point, NY 10996}
\affil[b]{U.S. Army Research Laboratory, Aberdeen Proving Ground, MD 21005}
\authorinfo{Further author information: (Send correspondence to Venkat R. Dasari)\\Venkat R.~Dasari: E-mail: venkateswara.r.dasari.civ@mail.mil, Telephone: +01-410-278-2846\\  Mee Seong Im: E-mail: meeseong.im@usma.edu, Telephone: +01-845-938-7461}

\titleformat{\paragraph}
{\normalfont\normalsize\bfseries}{\theparagraph}{1em}{}
\titlespacing*{\paragraph}
{0pt}{3.25ex plus 1ex minus .2ex}{1.5ex plus .2ex}

\begin{document} 
\maketitle

\begin{abstract}

Quantum applications  transmit and receive data  through quantum and classical communication channels. Channel capacity, the distance and the photon path between transmitting and receiving parties and the speed of the computation links play an essential role in timely synchronization and delivery of  information using classical and quantum channels. In this study, we analyze and optimize the parameters of the communication channels needed for the quantum application to successfully operate. We  also develop algorithms for synchronizing data delivery on classical and quantum channels. 

\end{abstract}

\section{Introduction}
\label{sec:intro}  

A programmable quantum network yields  remarkably flexible framework  to support quantum applications and abstract underlying hardware implementation details from the applications. The framework uses  quantum and classical channels as both are used by quantum applications to exchange information. Classical channels operate as  control channels to transmit metadata and measurement data between two nodes. Classical channels  are also used in error correction, diagnostics, and synchronization of communication. 
The flexibility may be understood as exponential speed and immense computing power which are critical for applications to absolute secure communication, interactive computation and infrastructure protection. These observables and new attributes at the quantum level can be controlled using  programmable network paradigm.

Optical switches are utilized in preserving the quantum states up to a unitary rotation $\hat{\sigma}_x$, $\hat{\sigma}_y$, or $\hat{\sigma}_z$, which are low loss, non-interferometric, and have potential to switch photon paths in less than $10^{-7}$ seconds. One method to enable superdense coding is Hong-Ou-Mandel interference by utilizing Bell states, requiring  $10^{-12}$ second timing precision between photon pairs during a Bell state measurement. However, transporting over large distances challenges this requirement due to difficulties with temperature that may be path-dependent with infrastructure dependencies, such as the distance and photon path between transmitting and receiving parties and the speed of the computation links.  In order to overcome physical interference problems, there is a need for optimization and  synchronization of quantum and classical channels. In this paper, we  study the optimization and synchronization related problems and propose  a set of models and optimization algorithms for the development of a stable  programmable quantum network for reliable operation of quantum applications.

\section{Related work and challenges}\label{section:related-work}
 
Optimization of quantum circuits is very important for creating reliable quantum networks, and there is an emerging body of research being carried out towards this goal\cite{nam2017automated, da2013global, perdomo2017readiness, armstrong2012programmable, furusawa2011quantum}. Programmable networks will provide a considerable function advantage and unforeseeable flexibility for quantum applications. Previously Openflow mediated  programmble quantum network model was proposed \cite{dasari2016openflow}, and simulation of programmable three node quantum  network model was demonstrated using mininet and CHP to  support superdense coding\cite{dasari2016programmable}. The fidelity of this design has been effectively validated through simulation and a numerical simulator.

We address developing, validating, and disseminating universal quantum network algorithms, filling in gaps to current technical procedures, and OpenFlow protocol is a key for the development, merging, and synchronization of classical and quantum channels using a remote administration, managing a network of forwarding elements. Data structures and communication functions are encoded using polarization-maintaining optical fibers and standard fiber optic cables, coherence-preserving switches (that induce multihop delays), and forwarding tables, further supporting the routing of quantum information. Using OpenFlow extensions for validation and to collect data, we solve timing issues arising from classical and quantum channels by providing several universal mathematical models, thus optimizing communication channel parameters with an algorithm to synchronize delivery of data through the two channels. Optimization of quantum circuits and classical channels,  in order to support quantum applications, has been a continuous effort by several research teams  across the world \cite{devetak2005capacity, cai2004quantum, hayashi2003general}. To elaborate further, quantum key distribution (QKD) is one of the widely studied quantum application and there are a set of stable protocols defined for successful implementation of QKD using quantum and classical networks \cite{barone1982physics}.  Generation of  cryptographic keys using QKD over controlled optical software defined network (SDN) was reported \cite{aguado2017secure}, and a comprehensive survey of software defined programmable networks is given in the paper by Nunes et al.\cite{nunes2014survey}

\section{Programmable Network Intelligence}

SDN is an emerging   and fast growing technology for interconnecting network devices and forwarding packets across them based on unified policies and security enforcements.  SDN promises deep programmability of network at all layers and it even extends the network state into applications for enabling them to make better pathing decisions. OpenFlow protocols play a key role in enabling SDN through its simple, flexible and adaptable architecture. OpenFlow provides a clean implementation of data and control plane separation that is essential to the success of SDN paradigm.  Originally OpenFlow is designed for working with packet switched networks and later versions of OpenFlow included support for circuit switched networks making it more comprehensive  in creating a converged control plane architecture for dynamical circuits and forwarding packets.  OpenFlow is even proved to be more efficient than generalized multiprotocol label switching (GMPLS) for managing optical networks.

In addition to providing forwarding intelligence and state distribution mechanisms, it also possesses topology discovery, collection of network performance statistics functions to make it a single most important protocol driving SDN architectures. Recent developments in quantum computational research have advanced our understanding of mechanisms of quantum phenomenon. However, there is a lack of progress and publications in defining mechanisms for secure quantum communications. Various ideas and models to propagate information related to quantum states are merely being postulated.

The standard test topology of a unified control plane consists of three layers. 
The software layer consists of host-specific applications 
necessary to access the quantum and classical resource on a node.  
Applications are not necessarily aware of the quantum communication channel, and this design emphasizes the ease of adapting existing software libraries to use over quantum communication. 

The middleware support abstraction translates the transmission and reception requests from the software layer into commands integrated by the quantum-enabled hardware. It captures knowledge about the range of uses acceptable to hardware and conform to an application programming interface. 

The hardware layer exposes access to the low-level physical resources within the node. It translates commands from the middleware layer into hardware specific commands for equipment, taking commands from the operating system and controlling the hardware for a device.

\subsection{Programmable framework to support quantum application}

The role of the classical channel manages the control data between the nodes and applications. Metadata such as the channel ID, application type, frequency and error correction codes control packets of information, which are essential in the advancement of a reliable communication.  Thus a programmable network is central to software-defined networking; it brings a network under software control such that packet-moving decisions are centralized, with network administrators remotely controlling and  allocating limited bandwidth and diverse traffic management resources.  We describe  mathematical models and algorithms to achieve the optimization and synchronization. We use QKD as a representative quantum application in describing our models.

Quantum physical layer is where the encoded quantum phenomenon occurs, which is not directly accessible by the classical layer due to physical restrictions. However, there is an indirect coordination between this layer and the classical layer, where the classical network traffic informs how each node and link should behave during the simulation. 
For example, when an entangled state is distributed across two nodes, its joint state depends on the local actions by each host application. The simulation server is called the dispatcher since it controls all classical interactions with the quantum layer. 
So the middleware connects directly to the dispatcher, emulating the physical link, centralizing quantum computation, and compiling observations using a quantum detector. 
Thus the phenomenon of quantum entanglement, quantum key distribution, superdense coding, and teleportation, all of which are crucial for quantum communication, occurs in this layer. 

Since the dispatcher has a holistic view of the quantum network, we control the dispatcher to model channel noise that is time-dependent and require bookkeeping of the order of transmissions on each quantum channel. 
The ARL quantum layer uses polarization-maintaining fibers (PMF) sending packetized entangled photons through its quantum programmable system. 

\section{Channel Synchronization and Optimization}

In most cases, both optical and classical channels use optical media for transmission; in some cases,  they were demonstrated to use free space  as well. Quantum channel paths are  directly connected where as classical channels can multihop due to circuit and packet switching, so the classical channel depends upon the network topology which results in a longer delay  for the signal to reach its destination. The proposed  programmable quantum communication framework (see Figures 1 and 2) was successfully simulated  using  mininet and CHP. However, due to the  architectural differences and the different paths the quantum and classical channels take,  there is a need  for optimization and synchronization of the communication channels in order to  reliably support quantum applications.





\subsection{Local decision algorithm at node level}\label{subsection:local-decision}

Node level local decision-making software module is critical in making a  determination of the synchronization state of quantum and classical channels. This  module  periodically obtains metadata from classical  and  quantum interfaces to determine the synchronization state of the communications.

Based on Algorithm $1$ below, we give three feasible, theoretical models in Sections~\ref{subsubsection:simple-model},  
\ref{paragraph:modified-PMF}, and \ref{paragraph:model-modified-multihop-delays}. 
Tensile and twisting stress, micro bending of the wires, and temperature and moisture of the environment are assumed to have negligible affect, and photons are of classical spherical shape, and reflection is assumed to retain elasticity.

\begin{figure}[htbp]\label{figure:qpn}
\centerline{\includegraphics[scale=.50]{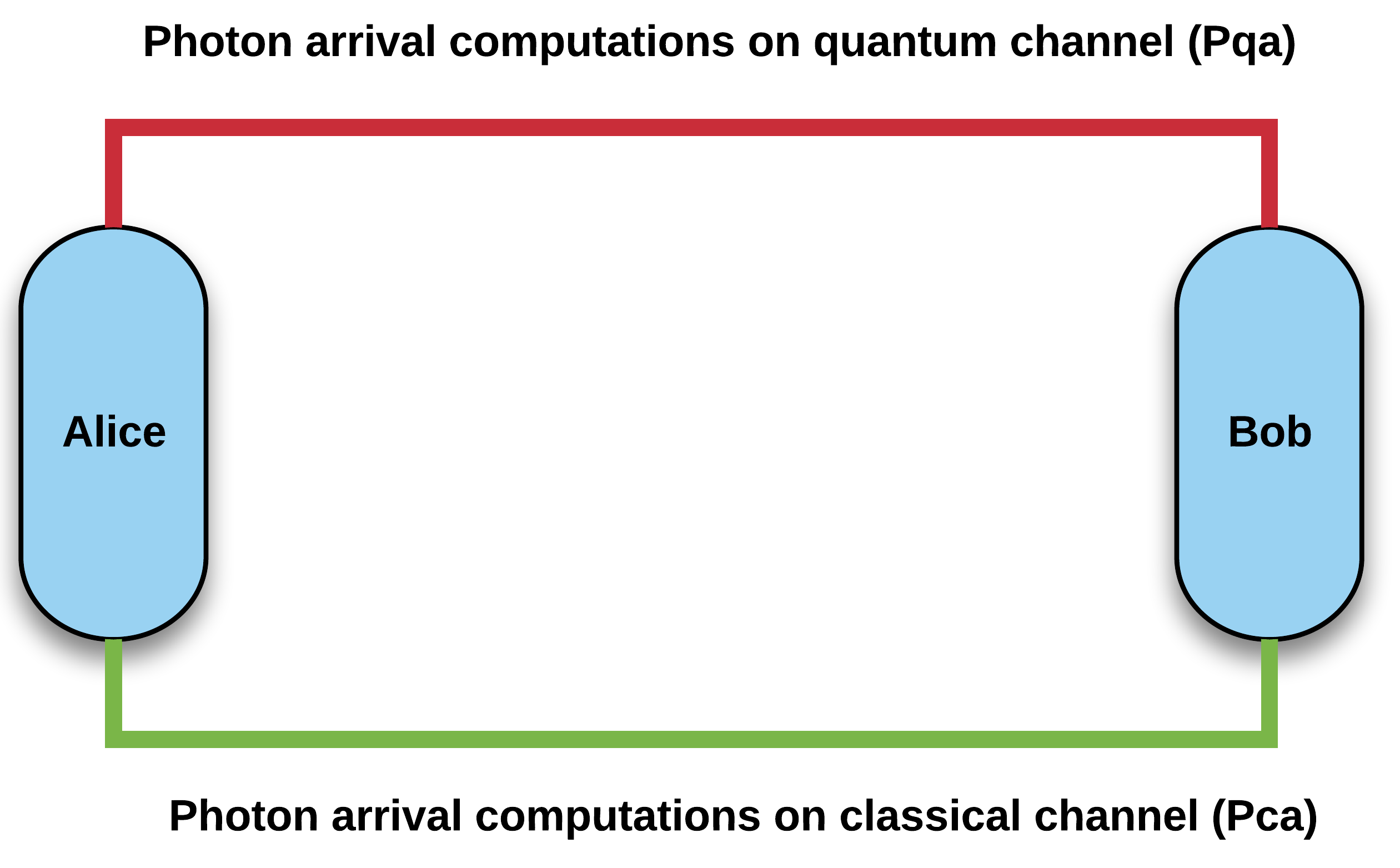}}
\caption{Channel synchronization scheme.}
\label{fig_qsimulator}
\end{figure}

\subsubsection{A linear model}\label{subsubsection:simple-model} 
We refer to Figure 1 for an illustration of our first model. A photon in a standard PMF is assumed to travel with an average velocity of $v_p=c/n_p$, where $c$ is the speed of light in a vacuum and $n_p$ is the index of refraction for the PMF (associated to our quantum channel), $1< n_p<3/2$.
 In a fiber optic cable made up of silica glass, the average speed of light through this medium is $v_f= (2/3)c$.  
 
Now assume that the multihop (propagation, processing, transmission, and queueing) delay is generic, with our model encompassing $k$ delays in series, each requiring $d$ amount of time per delay.  
Then the average distance a photon traveled in time $t$ in a PMF is $m_p(t) = tv_p$ while the average distance a photon traveled in a fiber optic cable is $m_f(t)=(t-kd)v_f$. 
In order to synchronize both channels, the quantum channel must have PMF of length $m_p(t)$ while the classical channel must have an optical wire of length $m_f(t)$. 

Now, in order to optimize both channels, packets of photons in the classical channel must arrive at a faster rate than the packets of photons in the quantum channel. One way to do this is by lengthening the PMF or shortening the classical optical wire. Since it is more efficient to shorten the classical wire with $k$ multihops, we will modify the optical cable wire $m_f(t)$. 

Let $T_{qa}$ be the arrival time of a photon through the quantum channel whose PMF has length $m_p(T)$ while $T_{ca}$ is the arrival time of a photon through the classical channel whose optical fiber has length $m_f(T)$. So our initial assumption is that photons through both channels arrive simultaneously, i.e., $T= T_{qa}=T_{ca}$. 
Let $T_{\Delta}=T_{ca}-T_{qa}$, the difference of the arrival time of a photon through the classical channel and the quantum channel. 
If we want a packet of photons to arrive $|T_{\Delta}|$ seconds faster through the classical channel than in the quantum channel, i.e., $|T_{\Delta}|=-T_{\Delta}$, then the length $m_f(T)$ of the fiber optic cable should be replaced with $m_f(T) - |T_{\Delta}|v_f$, giving us the following model: if the length of the PMF through the quantum channel is 
$$
\text{length}_{qa}  = m_p(T) = Tv_p,  
$$
then the length of the fiber optic cable on the classical channel must be 
$$
\text{length}_{ca}(T_{\Delta}) 
	= m_f(T) - |T_{\Delta}|v_f 
	= (T - kd)v_f - |T_{\Delta}|v_f 
$$
so that photons arrive $|T_{\Delta}|$ seconds faster (on average) on the classical channel. This technique allows us to control the classical network, making it comparable to that of the quantum channel. 


\makeatletter
\def\BState{\State\hskip-\ALG@thistlm}
\makeatother
\begin{algorithm}\label{algorithm:1}
\caption{Channel Synchronization }\label{euclid}
\begin{algorithmic}[1] 
\Procedure{Synchronization Process}{}
\State $\textit{Tqa} \gets \text{Photon arrival time  on the quantum channel }$ 
\State $\textit{Tca}\gets \text{Photon arrival time on the classical channel }$
\State $\textit{Td}\gets \text{Time difference between quantum and classical channels}$
\BState \emph{top}:

Td = Tca - Tqa  
\BState \emph{loop}:
\If {$\textit{Td}  \leq  \textit{0}$}
continue
\Else  
\\Drop photons

\State \textbf{goto} \emph{loop}.
\State \textbf{close};
\EndIf

\State \textbf{goto} \emph{top}.
\EndProcedure
\end{algorithmic}
\end{algorithm}


\subsubsection{A multinode, linear model}\label{subsubsection:multinode-linear-model} 
Under the same assumptions as in the second paragraph in Section~\ref{subsection:local-decision}, we elaborate the model given in Section~\ref{subsubsection:simple-model}. We refer to Figure 2 for an illustration of the  model in this section.  

\begin{figure}[htbp]\label{figure:qpn}
\centerline{\includegraphics[scale=.3]{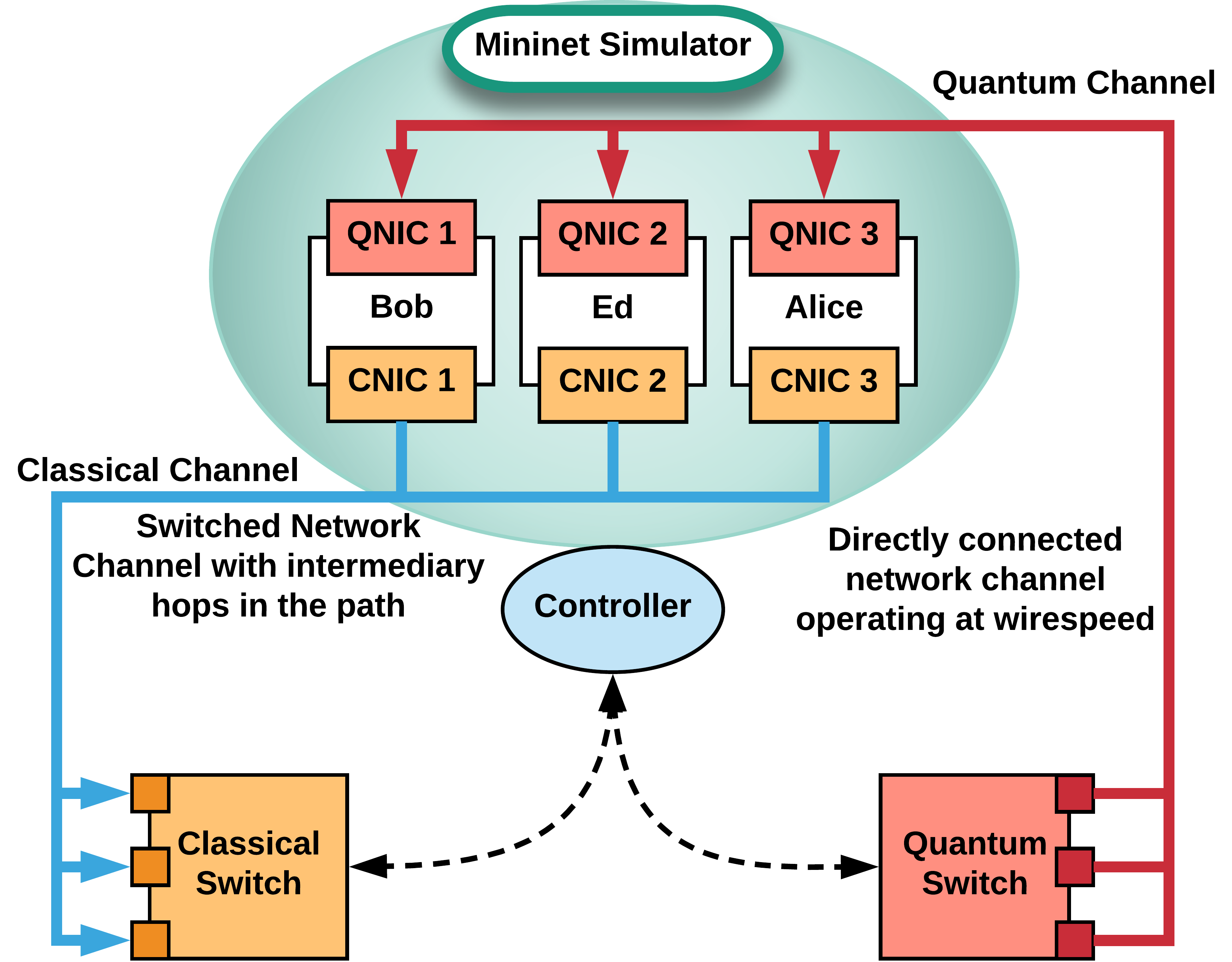}}
\caption{Programmable framework to support quantum application.}
\label{fig_qsimulator}
\end{figure}
 
Suppose one node is sending packets of photons via a centralized controller through both classical and quantum channels to $\ell>0$ nodes. For example, Bob wants to send a message to both Ed and Alice ($\ell=2$) through the quantum and classical channels. We want to synchronize quantum and classical channels for each node (note that the $\ell$ nodes need not be equidistant from the sender). Using the same notation as in Section~\ref{subsubsection:simple-model}, let $v_p=c/n_p$ be the average photon velocity through the PMF, where $1<n_p<3/2$ is the refraction index and $c$ is the speed of light in a vacuum.  
Assume that the average speed of light through a fiber optic cable is $v_f=(2/3)c$. 

Assume node $i$, $1\leq i\leq \ell$, has $k_i\geq 0$ delays in series, each requiring $d_{i,j}$ amount of time, 
$1\leq j\leq k_i$. Note that we do not need to assume that the delays for node $i$ are distinct from the delays for node $j$, where $j\not=i$. 
Then the average distance a photon traveled in time $t_i$ in a PMF to node $i$ is $m_p(t_i)=t_iv_p$ while the average distance a photon traveled in a fiber optics cable to node $i$ is $m_f^i(t_i)=(t_i-\sum_{j=1}^{k_i} d_{i,j})v_f$. 
In order to synchronize both channels at node $i$, the quantum channel must have PMF of length $m_p(t_i)$ while the classical fiber optic channel for node $i$ must have length $m_f^i(t_i)$. 

\paragraph{Modifying polarization-maintaining fibers in a multinode, linear model}\label{paragraph:modified-PMF}

Now suppose we want to optimize both channels at each node by controlling the arrival time of photons in the classical channel. Although it is more efficient to shorten the fiber optic cable, this may not always be possible. So we will modify and lengthen PMF, instead, in this model. 

Let $T_{qa}^i$ be the arrival time of a photon at node $i$ through the quantum channel whose PMF has length $m_p(T_i)$ and let $T_{ca}^i$ be the arrival time of a photon at node $i$ through the classical channel whose optical fiber has length $m_f^i(T_i)$. Here, assume $T_i=T_{qa}^i=T_{ca}^i$. 
Let $T_{\Delta}^i= T_{ca}^i-T_{qa}^i$. 
Suppose we want photons to arrive $|T_{\Delta}^i|$ seconds faster to node $i$ through the classical channel than in the quantum channel. Then we need to replace the length $m_p(T_i)$ of the PMF with $m_p(T_i)+|T_{\Delta}^i|v_p$. Repeat this substitution recursively for each node to obtain that: if the length of the fiber optic cable with $k_i$ delays in series, each requiring $d_{i,j}$ amount of time, $1\leq j\leq k_i$, is 
$$
\text{length}_{ca}^i 
	= m_f^i(T_i) 
	= \left(T_i-\sum_{j=1}^{k_i} d_{i,j}\right)v_f, 
$$
then the corresponding length of the PMF should be 
$$
\text{length}_{qa}^i(T_{\Delta}^i) 
	= m_p(T_i)+|T_{\Delta}^i|v_p 
	= T_iv_p + |T_{\Delta}^i|v_p, 
$$
where $i$ runs over $1$ through $\ell$, so that packets of photons arrive at the $i$-th node $|T_{\Delta}^i|$ seconds faster over the classical channel than through the quantum channel.


\paragraph{Modifying multihop delays in a multinode, linear model}\label{paragraph:model-modified-multihop-delays}
In this section, we will modify the internal programmable network by inserting and controlling multihop delays rather than modifying the PMF (or the fiber optic cable). 
Recall from Section~\ref{subsubsection:multinode-linear-model} that the length of a PMF for a photon traveling in the quantum channel to node $i$ over time $T_i$ is $m_p(T_i)=T_iv_p$ while the length of a fiber optic cable for a photon traveling in the classical channel to node $i$ over time $T_i$ is $m_f^i(T_i)=(T_i-\sum_{j=1}^{k_i} d_{i,j})v_f$. Assume that $T_i>0$ is carefully chosen (it is the minimum of all such values) so that photons arrive through the quantum and the classical channel synchronously. 
 
Let  $T_{\Delta}^i = T_{ca}^i - T_{qa}^i$, where $T_{ca}^i$ is the arrival time of a photon at node $i$ through the classical channel and $T_{qa}^i$ is the arrival time of a photon at node $i$ through the quantum channel (so $T_{ca}^i=T_{qa}^i=T_i$). 
 
Now, suppose we want to control time delay at node $i$ by speeding up the classical channel by $|T_{\Delta}^i|$ seconds while fixing the assumptions for the quantum channel. Although a merging of two multihop delays with a one delay or a substitution of a delay with a more modern and efficient (propagation, processing, transmission, and queueing) delay is possible, we will assume that a removal or a replacement of any multihop delay is not allowed in this model.  
Instead, assume that every pair of distinct delays in the classical network is connected by a unique fiber optic cable, i.e., we have a complete graph consisting of delays. 
Then in order to speed up the classical channel, we need to replace the $k_i$ delays for node $i$ in $m_f^i(T_i)$ with 
$m_f^{i,\Delta(r_i)}(T_i)= (T_i-\sum_{\alpha=1}^{r_i} d_{\alpha}^i)v_f$, 
for some $r_i$, $d_{\alpha}^i\in \{ d_{1,1},d_{1,2},d_{1,3},\ldots, d_{\ell,k_{\ell}}\}$, and $d_{\alpha}^i$ pairwise distinct for a fixed $i$, 
so that the inequality $\sum_{j=1}^{k_i}d_{i,j}-\sum_{\alpha=1}^{r_i} d_{\alpha}^i \geq |T_{\Delta}^i|$ holds.  
\begin{center}
Let $R_i\geq 1$ such that the inequality 
$\sum_{j=1}^{k_i}d_{i,j}-\sum_{\alpha=1}^{R_i} d_{\alpha}^i \geq |T_{\Delta}^i|$ holds and the difference $\sum_{j=1}^{k_i}d_{i,j}-\sum_{\alpha=1}^{R_i} d_{\alpha}^i$ is minimal.  
\end{center}
Define $m_f^{i,\Delta}(T_i) := (T_i-\sum_{\alpha=1}^{R_i} d_{\alpha}^i) v_f$. 

Note that there may be more than one combination of delays that will produce our $R_i$. 

Now we assume that the internal classical network has been programmed so that a photon heading towards node $i$ goes through the delays mentioned in $m_f^{i,\Delta}(T_i)$ (the uniqueness of the set of delays mentioned here is not a crucial concern in this manuscript since the delays are generic). 
If the quantum network requires $T_{qa}^i$ amount of time for the arrival of a photon to node $i$ over the PMF length  
\[
\text{length}_{qa}^i = m_p(T_i) =T_iv_p, 
\] 
then the classical network for node $i$ will send the photon through $R_i$ delays that require a total of $\sum_{\alpha=1}^{R_i} d_{\alpha}^i$ amount of time. So while the length of the optical fiber remains at 
\[ 
\text{length}_{ca}^i = m_f^i(T_i) =  \left(T_i-\sum_{j=1}^{k_i} d_{i,j}\right)v_f, 
\] 
the photon will arrive at the $i$-th node approximately $|T_{\Delta}^i|$ seconds faster.

\section{Discussion}

Quantum channels can carry and transmit quantum information, which are built on the principles of quantum physics. Because of the capacity of superposition and the capability of quantum entanglement of photons, they are expected to process information rapidly with immense computing power. 
 
We have proposed several simple theoretical models and optimization algorithms for the development of a stable programmable quantum network for reliable operation of quantum applications. Our models encode multihops and communication functions to recognize polarization-maintaining optical fibers in a synchronized classical-quantum channel. 
We plan to consider data structures, coherence-preserving switches and categorical temporal alignment that support 	quantum communications in our forthcoming study.

\section{Acknowledgements}

This work is supported by a research collaboration between the U.S. Army Research Laboratory and U.S. Military Academy. 
MSI thanks the U.S. Army Research Laboratory for their hospitality and an exceptional working environment. MSI is partially supported by National Research Council Research Associateship Programs.

\bibliography{im-dasari} 
\bibliographystyle{spiebib} 

\end{document}